\pdfoutput=1
\documentclass[aps,prb,floatfix,superscriptaddress,reprint]{revtex4-2}
\usepackage{eurosym}
\usepackage{amsbsy}
\usepackage{array}
\usepackage{pifont}
\usepackage{latexsym,epsfig,graphicx}
\usepackage{dcolumn}
\usepackage{subfigure}
\usepackage{comment}
\usepackage{multirow}
\usepackage{color}
\usepackage{bm}
\usepackage{mathrsfs}
\usepackage{amssymb}
\usepackage{amsfonts}
\usepackage{amsmath}
\usepackage{xspace}
\usepackage{float}
\usepackage{epstopdf}
\usepackage{tabularx}
\usepackage{longtable}
\usepackage[colorlinks=true, letterpaper=true, pdfstartview=FitV, linkcolor=blue, citecolor=blue, urlcolor=blue]{hyperref}
\usepackage[normalem]{ulem}
\usepackage[version=4]{mhchem}
\usepackage{epstopdf}

\begin{document}
\title{Tailoring Corner States and Exceptional Points in Altermagnets}
\author{Xiao-Ming Zhao}
\email{xmzhao@ustb.edu.cn}
\affiliation{Department of Physics and Institute of Theoretical Physics, University of Science and Technology Beijing, Beijing 100083, China}
\affiliation{Key Laboratory of Multiscale Spin Physics (Ministry of Education), Beijing Normal University, Beijing 100875, China}
\author{Cui-Xian Guo}
\affiliation{Beijing Key Laboratory of Optical Detection Technology for Oil and Gas, China University of Petroleum-Beijing, Beijing 102249, China}
\affiliation{Basic Research Center for Energy Inter disciplinary, College of Science, China University of Petroleum-Beijing, Beijing 102249, China}
\author{Xin-Ran Ma}
\affiliation{Center for Advanced Quantum Studies, Department of Physics, Beijing Normal University, Beijing 100875, China}
\author{Xiao-Ran Wang}
\affiliation{College of Teacher Education, Hebei Normal University, Shijiazhuang 050024, China}
\author{Su-Peng Kou}
\email{spkou@bnu.edu.cn}
\affiliation{Center for Advanced Quantum Studies, Department of Physics, Beijing Normal University, Beijing 100875, China}

\begin{abstract}
Altermagnets (AMs) exhibit vanishing net magnetization but strong momentum-dependent spin splitting enforced by crystal symmetry. Here, we explore the non-Hermitian effects in dissipative two-dimensional AMs. We show that symmetry-compliant dissipation naturally induces an imaginary staggered exchange field, driving a NH topological phase transition absent in conventional antiferromagnets. In the topologically nontrivial phase, hybrid skin-topological modes driven by altermagnetic $d$-wave anisotropy emerge, as captured by the chiral skin effect framework. In the gapless phase, we elucidate the creation and annihilation dynamics of exceptional points. Crucially, we analytically prove via the transfer matrix method that corner states are deterministically controlled by the boundary sublattice termination. Owing to the symmetry constraints and the robustness of  chiral states, these findings hold universally across all topological AMs. A general framework is established for controlling topological corner states, offering a new strategy for designing magnetic materials with tailored non-Hermitian properties.
\end{abstract}
\maketitle

\emph{\textbf{Introduction.}}---Altermagnetism has recently emerged as a third fundamental magnetic phase, distinct from conventional ferromagnetism and antiferromagnetism \cite{AM1,AM2,AMAdd1,AM3,AM4,AM5,AM6,AM7,AM8,AM9,AM10,AM11,AM12,AM13}.  Characterized by vanishing net magnetization yet exhibiting strong, symmetry-enforced momentum-dependent spin splitting, altermagnets (AMs) host nonrelativistic $d$-wave nodal structures in momentum space originating from crystal-symmetry-mediated exchange coupling rather than relativistic spin-orbit coupling (SOC) \cite{AM2,AM9}. This unique electronic structure gives rise to a range of distinctive phenomena, including the anomalous Hall and Nernst effects \cite{MacNH2,MacNH2.1,MacNH2.2,MacNH2.3,MacNH2.4}, the magneto-optical Kerr effect \cite{ker1,ker2,ker3}, magnetoresistance \cite{TunMag1,TunMag2,TunMag3,TunMag4}, nonrelativistic spin–charge conversion \cite{ChaSpin1,ChaSpin2,ChaSpin3,ChaSpin4,ChaSpin6,ChaSpin7}, and unconventional piezomagnetism \cite{UncPie1,UncPie2,UncPie3,UncPie4}. These properties not only differentiate AMs from other magnetic phases but also underscore their potential for applications in spintronics \cite{Apply1,Apply2,Apply3,Apply4,Apply5} and nonvolatile memory devices \cite{MMemery1,MMemery2}.

However, realistic magnetic systems are inevitably coupled to their environments, leading to intrinsic dynamical damping, magnon decay, or finite quasiparticle lifetimes, which endow these systems with a natural predisposition toward non-Hermitian (NH) physics \cite{MacNH1,MacNH2,MacNH3,MacNH4,MacNH5}. While the NH effects have been extensively explored in various platforms, its interplay with the unique symmetry landscape of AMs remains an open frontier. A fundamental question arises: How does the momentum-dependent spin texture of AMs imprint itself onto dissipative processes? In conventional antiferromagnets (AFMs), dissipation typically results in isotropic broadening, yielding only trivial complex energy shifts. In stark contrast, the symmetry-enforced nodal structure of AMs suggests that dissipation must also be highly anisotropic, potentially driving novel NH topological phases that have no counterpart in standard magnetic systems.

In this work, we address this challenge by investigating the NH topology of two-dimensional (2D) AMs subject to symmetry-compliant dissipation. We derive a microscopic effective model demonstrating that the imaginary staggered exchange field arises naturally from a dissipation-induced self-energy, which is `locked' to the altermagnetic spin texture. We show that this interplay drives a NH topological phase transition, distinct from the AFM limit. In the topologically nontrivial phase, we report the emergence of hybrid topological-skin effect. These robust boundary modes are driven by the altermagnetic $d$-wave anisotropy, which induces direction-dependent decay rates and dissipative refraction at the corners. Furthermore, we elucidate the creation and annihilation dynamics of exceptional points (EPs) in the gapless regime. Crucially, employing a transfer matrix approach, we derive an analytical criterion proving that the spatial localization of corner states is deterministically controlled by the boundary sublattice termination, mechanistically driven by the NH chiral skin effect. Our findings establish a general framework for tailoring robust corner states in magnetic materials through boundary engineering.

\clearpage
\begin{figure}[tbp]
\centering
\includegraphics[clip,width=0.45\textwidth]{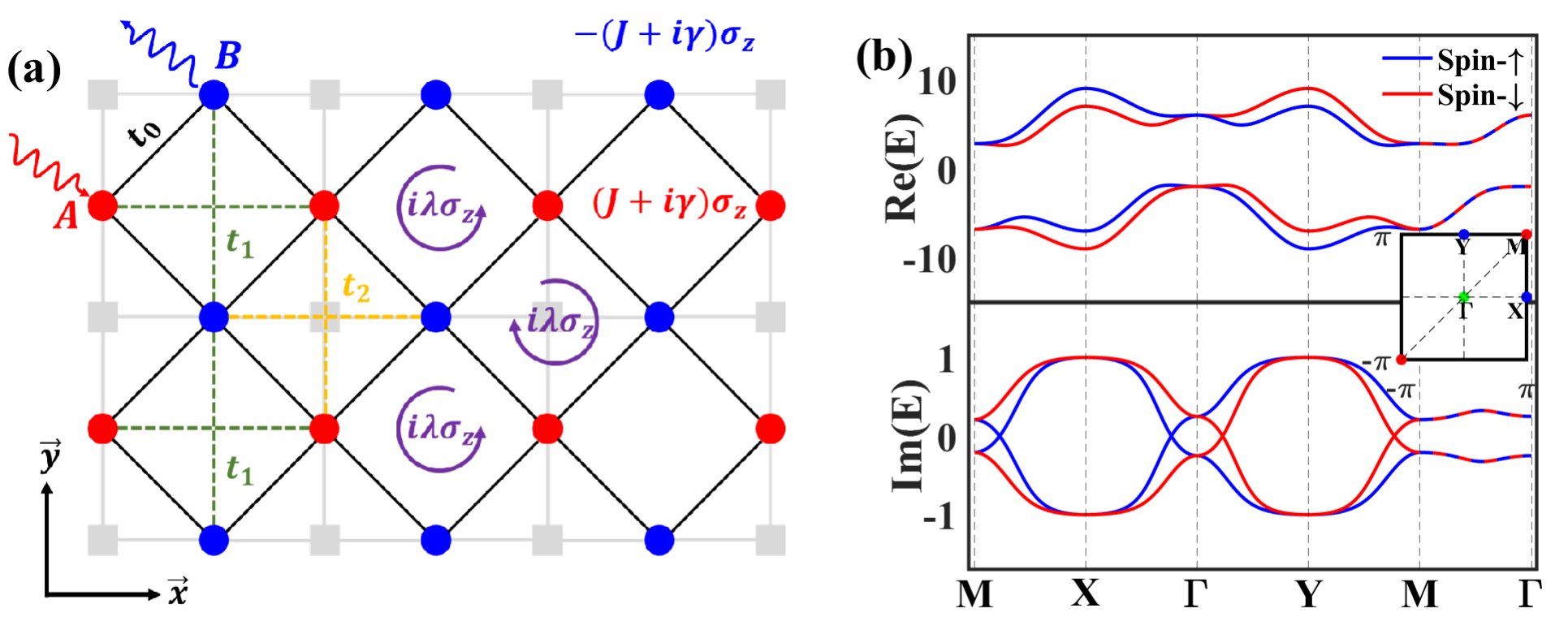}
\caption{Non-Hermitian altermagnet model and band structure.
(a) Lattice schematic where staggered dissipation ($\pm i\gamma$) is strictly locked to magnetic sublattices A (red) and B (blue) with opposite Néel vectors. Altermagnetic anisotropy is enforced by unequal nearest-neighbor hoppings $t_1 \neq t_2$.
(b) Typical complex band structure showing characteristic spin-splitting, which solely from the anisotropy and vanishes in the isotropic AFM limit ($t_1=t_2$). Parameters used: $t_{1}=0.5$, $t_{2}=2$, $J=1$, $\lambda=1.2$, $\gamma=1$.}
\label{Hamiltonian}
\end{figure}

\emph{\textbf{Model and symmetry analysis.}}---To investigate the NH effects in dissipative AMs, we consider a 2D $d_{xy}$ AM on a lieb-type lattice, as depicted in Fig.~\ref{Hamiltonian}(a), which comprises two magnetic sublattices and one nonmagnetic ligand site. Rather than introducing non-Hermiticity phenomenologically, we derive our effective model from a microscopic open quantum system \cite{Supplemental}. We consider itinerant electrons coupled to the local Néel order $\mathbf{S}_\mu$ ($\mu \in \{A, B\}$) via $sd$-hybridization, while simultaneously interacting with a dissipative bath. Integrating out the bath degrees of freedom yields a spin-dependent self-energy correction:
\begin{eqnarray}\label{SelfEnerg}
\Sigma_\mu \approx -i\gamma (\mathbf{S}_\mu \cdot \boldsymbol{\sigma}),
\end{eqnarray}
where $\gamma$ characterizes the effective coupling strength to the bath. This relation is pivotal: it demonstrates that the dissipation is not a uniform background but is strictly `locked' to the local magnetic texture.

By projecting out the nonmagnetic ligand sites \cite{MinModel1,MinModel2,MinModel3,MinModel4}, we obtain the effective Hamiltonian for the itinerant electrons : 
\begin{eqnarray}\label{Ham}
H(\textbf{k})=H_{0}+H_{SOC}+H_{AM}
\end{eqnarray}
where $H_{0}=(h_{0}\tau_{0}+h_{1}\tau_{x}+h_{3}\tau_{z})\sigma_{0}$ describes the hopping kinetics with $h_{0}(\mathbf{k})=-t_{a}(\cos k_{x}+ \cos k_{y}),
h_{1}(\mathbf{k})=-4t_{0}\cos(\frac{k_{x}}{2})\cos(\frac{k_{y}}{2}),
h_{2}(\mathbf{k})=4\lambda \sin(\frac{k_{x}}{2})\sin(\frac{k_{y}}{2}),
h_{3}(\mathbf{k})=-t_{b}(\cos k_{x}-\cos k_{y})$, with nearest-neighbor hopping $t_0=1$ between the A and B sublattices, and anisotropic second-nearest-neighbor hoppings $t_1=(t_{a}+t_{b})/2$ and $t_2=(t_{a}-t_{b})/2$. $H_{SOC}=h_{2}\tau_{y}\sigma_{z}$ represents the spin-orbit coupling. Here $\tau$ and $\sigma$ are Pauli matrices in the sublattice and spin spaces, respectively. The hallmark of our model is the complex altermagnetic term:
\begin{eqnarray}
H_{AM}=(J+i\gamma )\tau_{z}\sigma_{z}
\end{eqnarray}
Here, the real part $J$ arises from the conventional $sd$-exchange interaction, while the imaginary part $i\gamma$ represents the dissipation-induced lifetime correction. Crucially, this NH term is strictly proportional to the staggered magnetization operator $\tau_z\sigma_z$. This reflects a `locking' mechanism: since the system–bath coupling is mediated by the local magnetic moments, the resulting dissipation inherits the staggered spatial symmetry of the Néel order \cite{Supplemental}.

\begin{figure}[tbp]
\centering
\includegraphics[clip,width=0.28\textwidth]{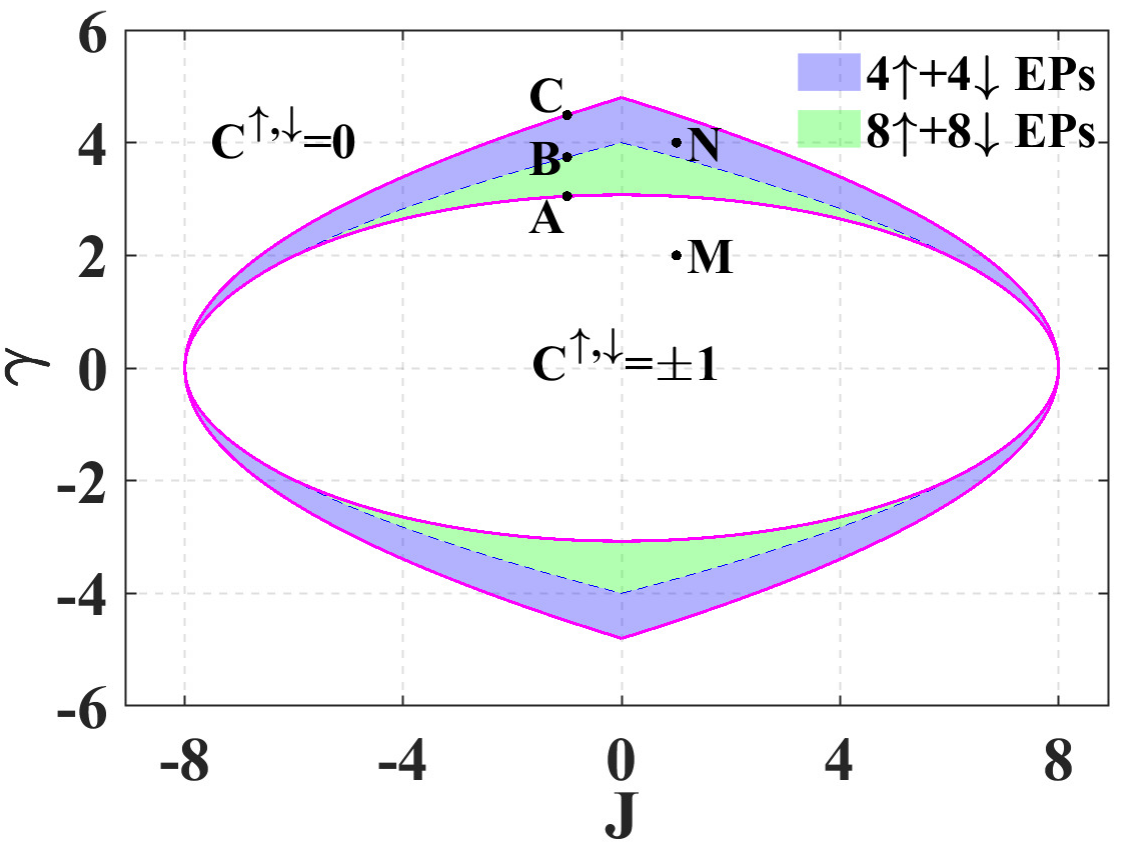}
\caption{Phase diagram induced by the complex exchange field. White regions distinguish the central topological Chern insulator (point M) from the outer trivial phase. Colored regions denote gapless phases with exceptional points. Parameters: $t_{1}=0.5$, $t_{2}=2$, $\lambda=1.2$, $\gamma=2$. This rich topology vanishes completely in the conventional AFM limit, where the entire diagram is trivial.}
\label{PhaseDiag}
\end{figure}

The spectral properties and topological features of the system can be systematically elucidated through a symmetry-based analysis. Initially, the Dirac points in $H_{0}$ are protected by mirror symmetries $M_{x/y}$, yet the introduction of SOC breaks these symmetries by altering the spin texture, thereby opening a band gap and facilitating band inversion—a crucial step toward the formation of topological boundary states. Meanwhile, the system maintains a horizontal mirror symmetry $M_{z}$($z$$\rightarrow$$-z$). This allows the Hamiltonian matrix to be decoupled into two independent blocks, each corresponding to a distinct spin channel. As a result, the full Hamiltonian can then be written as $H(\textbf{k})=[H^{\uparrow}_{\textbf{k}},0;0,H^{\downarrow}_{\textbf{k}}]$. The energy spectrum is
\begin{eqnarray}
E_{\pm}=h_{0}\pm\sqrt{h_{1}^{2}+h_{2}^{2}+[h_{3}+\epsilon(J+i\gamma)]^{2}}
\end{eqnarray}
 with $\epsilon=\pm1$ for spin $\uparrow$/$\downarrow$. In the AFM limit ($t_b=0$), $h_3$ vanishes, reducing the system to a conventional square-lattice AFM with spin degeneracy. In contrast, a finite $t_{b}$ enforces a $d$-wave modulation, lifting the degeneracy almost universally except along the diagonal $k_{x}=\pm k_{y}$ (line $\Gamma$-$M$ in Brillouin zone (BZ)), yielding a $d_{x^{2}-y^{2}}$ symmetric Zeeman splitting. A rigorous analytical comparison between the AM phase and the AFM limit is provided in Supplemental Material Sec. IV. As we demonstrate below, this symmetry-enforced anisotropy is the fundamental resource that enables both the bulk NH topology and the direction-dependent decay rates essential for the chiral skin effect. Finally, the combined fourfold rotation symmetry $C_{4z}$  and time-reversal symmetry $T$ —often termed symmetry-enforced spin-momentum locking—permitting momentum-dependent alternating spin splitting as clearly visible in Fig.~\ref{Hamiltonian}(b). Crucially, this splitting stems from the collinear magnetic order rather than relativistic SOC.

\emph{\textbf{NH topological phase transition}}---The interplay between the altermagnetic anisotropy and the symmetry-compliant dissipation drives a rich topological phase diagram, as mapped in Fig.~\ref{PhaseDiag}. The parameter space is partitioned into gapped topological phases (white regions containing point M), trivial insulating phases (outer white regions), and gapless phases hosting EPs (colored regions), where eigenvalues and eigenstates coalesce \cite{EP1,EP2,EP3,EP4}.

\begin{figure}[tbp]
\centering
\includegraphics[clip,width=0.49\textwidth]{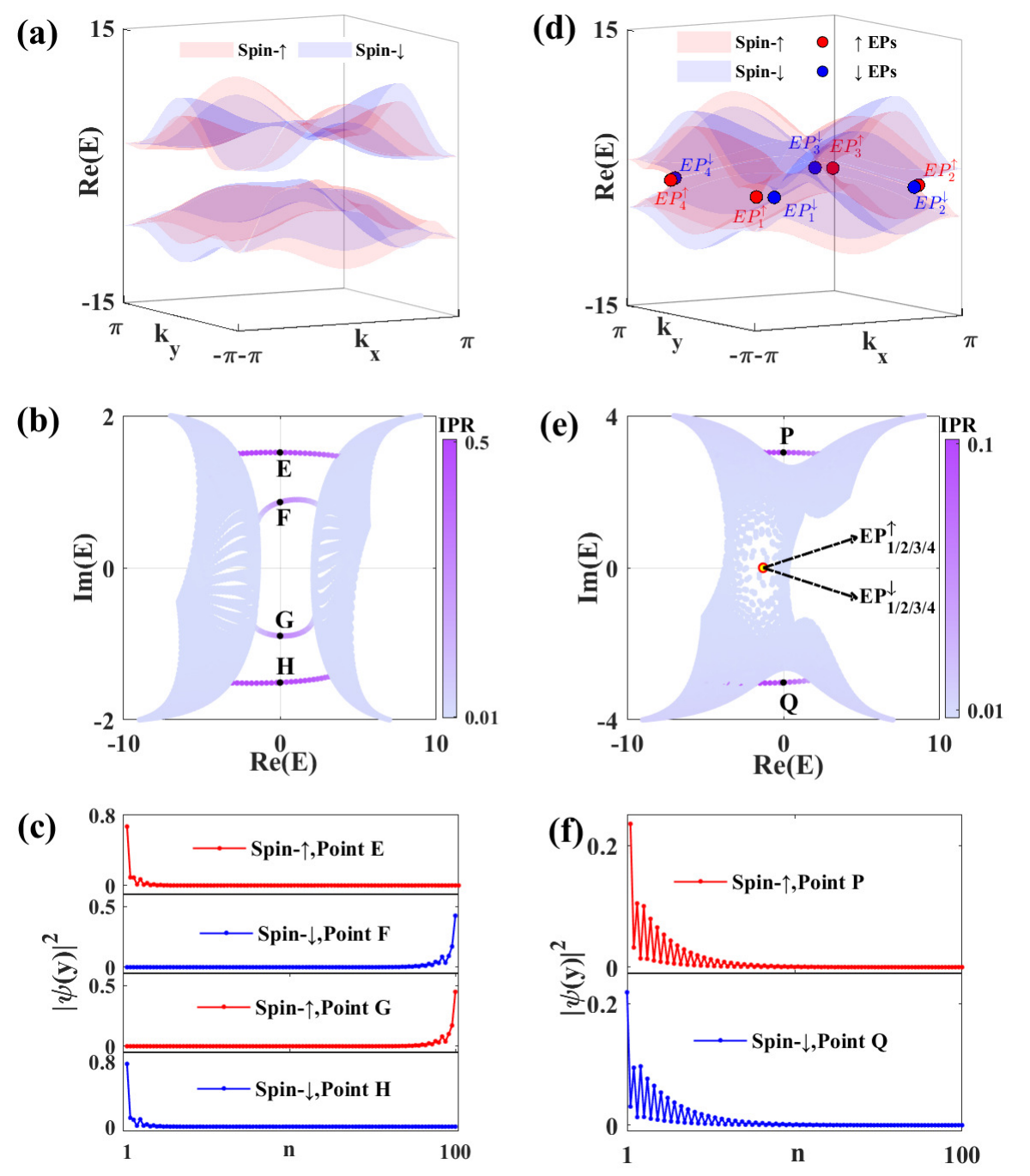}
\caption{ Energy spectra and boundary modes in topological and gapless phase. \textbf{(a–c)} Results for the gapped topological phase (point M in Fig.~\ref{PhaseDiag}): (a) Real part of the complex energy spectrum under $x,y$-PBC, indicating a clear bulk gap. (b) Complex energy spectrum under $x$-PBC/$y$-OBC, where darker color denote in-gap helical edge modes. (c) Spatial distribution of the edge modes corresponding to points $E$, $G$, $F$ and $H$ in (b). \textbf{(d–f)} Analogous results for the gapless phase (point N), where each spin sector hosts four EPs. States near points $P$ and $Q$ in (e) exhibit boundary localization at $y=1$, as illustrated in (f). Crucially, these modes are topologically unprotected.}
\label{TopEdgeState}
\end{figure}

It is essential to highlight the unique role of altermagnetism here. In the AFM limit ($t_b=0$), the dispersive mass term $h_3(\boldsymbol{k})$ vanishes, leaving only a momentum-independent effective mass $\epsilon J + i\gamma$. Consequently, the band inversion required for topological transitions is precluded \cite{Supplemental}, rendering the system topologically trivial regardless of the dissipation strength. Thus, the momentum-dependent spin splitting intrinsic to AMs is not merely a background but the fundamental prerequisite for the realization of the NH topological phases.

To characterize the topology in the NH regime, the Chern number must be formulated within the biorthogonal basis to properly account for the non-orthogonality of right ($|u_R\rangle$) and left ($|u_L\rangle$) eigenstates. The spin-resolved Chern numbers are defined as:
\begin{eqnarray}
C^{\sigma} = \frac{i}{2\pi}\int_{BZ} \epsilon_{ij} \langle \partial_{k_i} u_L^\sigma | \partial_{k_j} u_R^\sigma \rangle d^2k, 
\end{eqnarray}
where $\epsilon_{ij}$ is the antisymmetric Levi-Civita tensor ($i,j \in \{x,y\}$) \cite{ChenNH} and summation over repeated indices is implied. A non-zero $C^{\sigma}$ dictates the emergence of chiral edge states under open boundary conditions (OBC), governed by the NH bulk-boundary correspondence. 

Within the topological phase (e.g., at Point M), the real part of the spectrum exhibits a clear bulk gap, as shown in Fig.~\ref{TopEdgeState}(a). Under strip geometry ($x$-PBC, $y$-OBC), gapless chiral boundary modes emerge within the gap. These modes are spatially localized at the edges, as visualized by the Inverse Participation Ratio (IPR) in Fig.~\ref{TopEdgeState}(b), defined as $\text{IPR}(k_{x})=\sum_{y}|u_{R}|^{4}/(\sum_{y}|u_{R}|^{2})^{2}$. While darker shades in Fig.~\ref{TopEdgeState}(b) indicate strong localization, the wave-function profiles in Fig.~\ref{TopEdgeState}(c) explicitly confirm that these states are spatially confined to the system ends.

\begin{figure}[tbp]
\centering
\includegraphics[clip,width=0.48\textwidth]{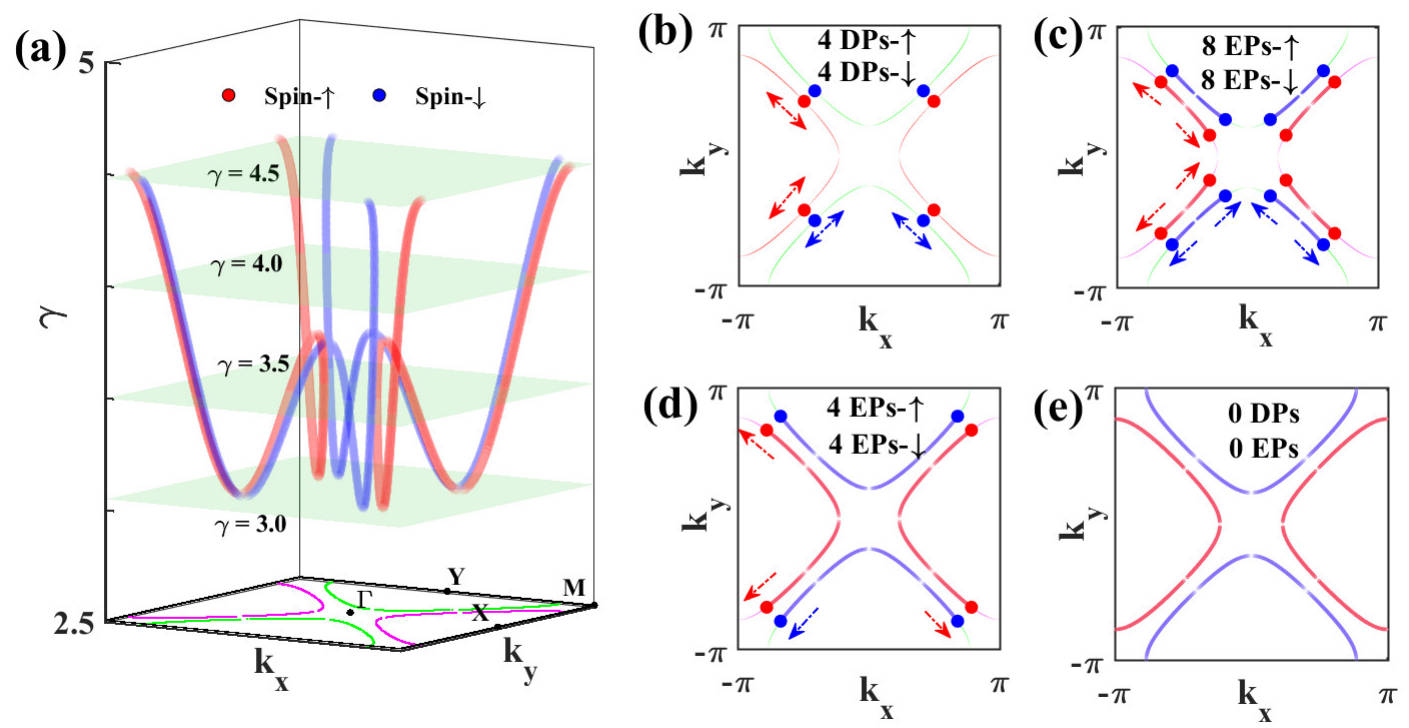}
\caption{Symmetry-protected evolution of EPs in momentum space with increasing NH strength $\gamma$.
(a)Continuous trajectories of spin-up (red) and spin-down (blue) EPs in the $k_x$–$k_y$ plane as a function of $\gamma$, their projections onto the basal plane of the basal plane are indicated by pink and green curves. Crucially, the two sets of trajectories are distinct by a $\pi/2$ rotation, directly reflecting the altermagnetic $C_{4z}$ symmetry. 
(b)–(e) Snapshots of precise EP positions at $\gamma=$3, 3.5, 4, 4.5. Panels (b) and (e) correspond to points A and C in the phase diagram of Fig.~\ref{PhaseDiag}, respectively. The EP locations correspond to intersections between the light-green horizontal plane and the trajectories in (a). Arrows indicate the direction of EP emergence and annihilation dynamics. Parameters: $\lambda = 1.2$, $t_1 = 0.5$, $t_2 = 2$, $J = 1$.}
\label{EPEvol}
\end{figure}

\begin{figure*}[tbp]
\centering
\includegraphics[clip,width=0.9\textwidth]{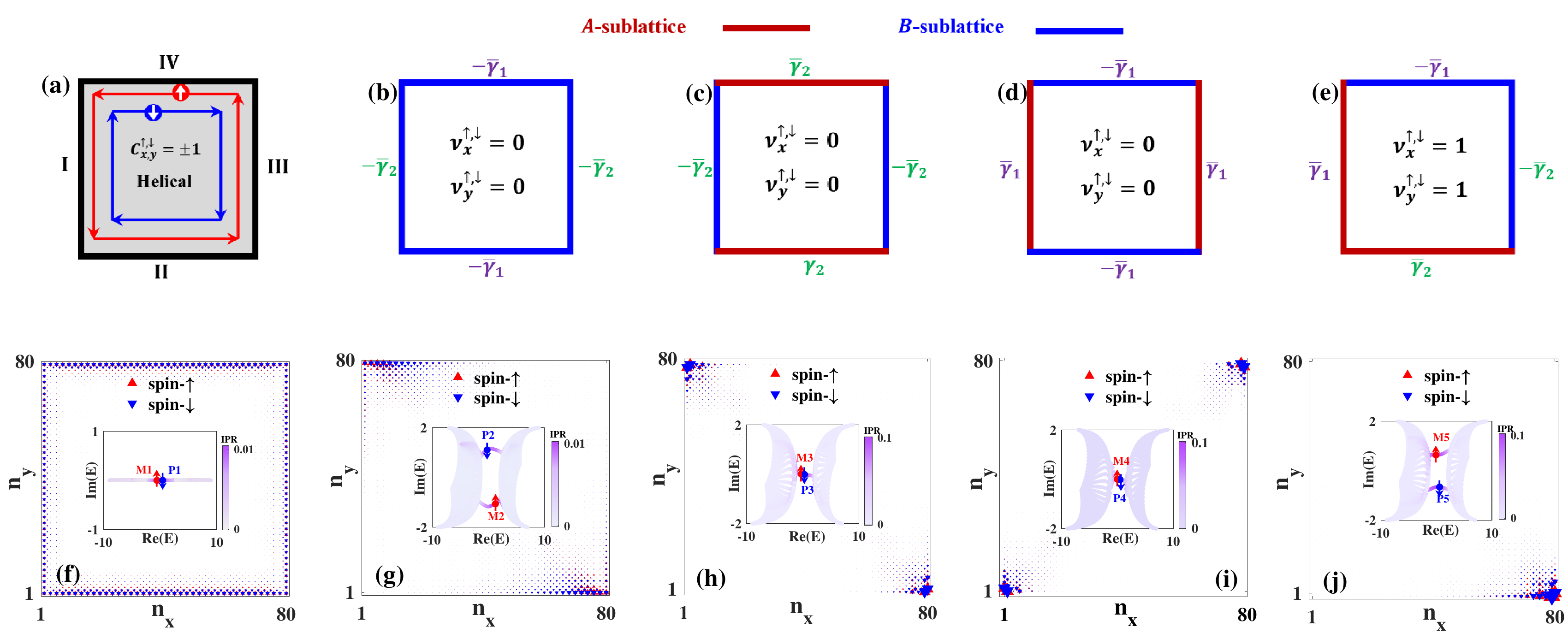}
\caption{Deterministic control of hybrid skin-topological corner modes via edge terminations. (a) Schematic showing helical edge states on edges I–IV. (b–e) Four distinct edge configurations (BBBB, BABA, ABAB, AABB) defined by the outermost sublattice (red A/blue B), illustrating the direction-dependent effective dissipation with $\overline{\gamma}_{1}\simeq0.4\gamma\sigma_{z}$ and $\overline{\gamma}_{2}\simeq0.7\gamma\sigma_{z}$. (f–j) Spatial profiles under $x,y$-OBC; insets show IPR-colored spectra.
(f) The Hermitian limit ($\gamma=0$), exhibiting uniform edge states that are independent of boundary configuration.
(g–j) The non-Hermitian case ($\gamma=2$, point M in Fig.~\ref{PhaseDiag}(b)). The corner localization shifts deterministically due to altermagnetic anisotropy-induced dissipative potentials. Crucially, this effect vanishes completely in the conventional AFM limit ($t_b=0$). Parameters: $\lambda=1.2$, $J=1$.}
\label{CornerState}
\end{figure*}

In the gapless regions (e.g., Point N), the interplay between the dispersive altermagnetic term $h_3(\boldsymbol{k})$ and dissipation $i\gamma$ generates EPs satisfying the coalescence conditions
\begin{eqnarray}\label{EPsCond}
h_{3}+\epsilon J=0, \text{   }\text{   }\text{   }h_{1}^{2}+h_{2}^{2}=\gamma^{2}.
\end{eqnarray}
Furthermore, the emergence of these robust EPs relies fundamentally on the altermagnetic symmetry; In the isotropic AFM limit, the requisite momentum-dependent mass term vanishes, strictly precluding the formation of EPs \cite{Supplemental}. We trace their dynamic evolution along the path A$\to$B$\to$C in Fig.~\ref{PhaseDiag}: initially, eight EPs emerge per spin sector, which subsequently annihilate in pairs upon increasing $\gamma$, as detailed in Fig.~\ref{EPEvol}. This evolution is strictly constrained by symmetry, with $C_{4z}$ dictating a $\pi/2$ rotation between spin-$\uparrow$ and spin-$\downarrow$ EP distributions. Notably, while boundary states still form in this regime [Fig.~\ref{TopEdgeState}(e,f)], they lack topological protection, contrasting sharply with the robust chiral corner modes observed in the gapped phase. 

\emph{\textbf{Hybrid skin-topological effect}}---Driven by the altermagnetic crystal symmetry, the spatial localization of chiral boundary modes manifests as a unique boundary phenomenon, while the bulk states remain extended. We identify this as the hybrid skin-topological effect \cite{NHSE1,NHSE2,NHSE3,NHSE4,NHSE5,NHSE6,NHSE7,NHSE8,NHSE9,
NHSE10,NHSE11,NHSE12,NHSE13,NHSE14,NHSE15,NHSE16,NHSE17,NHSE18,NHSE19,NHSE20,NHSE21,NHSE22,NHSE23,
NHSE24,NHSE25,NHSE26,NHSEadd1,NHSEadd2,NHSEadd3,NHSEadd4,Re1}, characterized by corner modes that scale linearly with system size [$\mathcal{O}(L)$] (see Fig.~\ref{TopEdgeState}(b)). This scaling behavior fundamentally distinguishes them from conventional second-order topological corner modes [$\mathcal{O}(1)$] and first-order NH skin modes [$\mathcal{O}(L^2)$], representing a distinct interplay between topological protection and NH localization \cite{NHTop1,NHTop2,NHTop3,NHTop4,NHTop5,NHTop6}.

A distinct feature of our approach is the derivation of an exact analytical solution for the NH edge spectra, a capability often absent in standard effective Hamiltonian methods. Using the transfer-matrix method \cite{TransMatrix}, we obtain the complex dispersion relation \cite{Supplemental}:
\begin{eqnarray}\label{EdgeEnergy}
E_{edge}(k_{i})=\alpha (k_{i})+i\gamma\beta(k_{i}),
\end{eqnarray}
where the real part $\alpha(k)$ recovers the Hermitian edge dispersion, while the imaginary part $i\gamma\beta(k)$ acts as an effective NH potential. This analytical form allows us to rigorously apply the chiral skin effect (CSE) theory to predict corner localization \cite{Re2}. According to the CSE framework, the accumulation of boundary states is determined by the interplay between the group velocity $v = \partial_k \alpha(k)$ and the spatial gradient of the global dissipation potential $\Gamma_g(r)$ (see section II.B of Ref.~\cite{Supplemental}). Specifically, states localize at a position $r_0$ (e.g., a corner) if and only if the potential creates a `trap' satisfying the criterion:
\begin{eqnarray}\label{CornState}
\Gamma_{g}(r_{0}-0^{+})>0,\Gamma_{g}(r_{0}+0^{+})<0.
\end{eqnarray}
This condition (derived as Eq. (\textcolor[rgb]{0.00,0.07,1.00}{S29}) in Ref.~\cite{Supplemental}) corresponds to a domain wall where the effective dissipation shifts from gain-like to loss-like along the direction of chiral propagation.

Applying this criterion explains the deterministic control observed in Fig.~\ref{CornerState}. Even with fixed bulk parameters, the edge termination (A or B sublattice) modulates the potential $\Gamma(r)$ \cite{Supplemental}. For instance, in the `BBBB' configuration [Fig.~\ref{CornerState}(b)], the anisotropy-induced mismatch $|\bar{\Gamma}_1 - \bar{\Gamma}_2|$ creates dissipative refraction points, trapping states at the Bottom-Right and Top-Left corners. Conversely, alternating configurations like `ABAB' [Fig.~\ref{CornerState}(d)] shift these domain walls, relocating the corner modes.

This behavior challenges previous theoretical frameworks relying on spectral point gaps \cite{ConerState} or non-zero winding numbers. As evidenced in Fig.~\ref{CornerState}, corner localization varies distinctly with sublattice termination even when the winding numbers vanish [$(\nu_x, \nu_y)=(0,0)$]. This confirms that the observed corner modes are not artifacts of isolated boundary sites but originate from the global chiral skin effect, a mechanism that intrinsically transcends local boundary descriptions \cite{Re1}.

\emph{\textbf{Conclusion.}}---In summary, we have established a microscopic framework for NH altermagnetism, demonstrating that symmetry-compliant dissipation inherently drives a topological phase transition distinct from conventional antiferromagnetism. By deriving the effective model from an open quantum system, we showed that the dissipation is `locked' to the magnetic texture. This interplay gives rise to two distinctive phenomena: the emergence of hybrid skin-topological corner modes at the boundaries, and the symmetry-protected creation and annihilation dynamics of EPs in the bulk.

These phenomena serve as unique fingerprints of NH altermagnetism, rooted in the symmetry-enforced $d$-wave anisotropy that is strictly absent in conventional AFMs. By inducing direction-dependent dissipative potentials, this anisotropy—combined with staggered boundary terminations—deterministically governs the spatial localization of corner modes.

Our results identify both the corner skin effect and the unique EP dynamics as defining fingerprints of NH altermagnetism, strictly absent in conventional AFMs. This work establishes a general roadmap for engineering robust corner states in synthetic platforms, such as superconducting circuits, cold-atom lattices, and acoustic metamaterials, opening new avenues for exploring dissipation-tailored spintronics.

\section*{DATA AVAILABILITY}
The data and source code that support the findings of this study are openly available in the GitHub repository at Ref.~\cite{Date}.

\begin{acknowledgments}
This work is supported by Guangdong Basic and Applied Basic Research Foundation (Grant No. 2023A1515110081), Open Fund of Key Laboratory of Multiscale Spin Physics (Ministry of Education), Beijing Normal University (Grant No. SPIN2024K01) and Fundamental Research Funds for the Central Universities, China (Grant No. FRF-TP-22-098A1), National Key R\&D Program of China (Grant No. 2023YFA1406704), National Natural ScienceFoundation of China (Grant Nos. 12174030,12405030).
\end{acknowledgments}

\end{document}